\pdfoutput = 1
\documentclass[aps,prb,reprint,showpacs,superscriptaddress]{revtex4-1}
\usepackage{graphicx}
\usepackage{graphics}
\usepackage{amsmath}
\usepackage{amssymb}
\usepackage{amsfonts}
\usepackage{dcolumn}
\usepackage{dsfont}
\usepackage{latexsym}
\usepackage{rotating}
\usepackage{array}
\usepackage{makecell}
\usepackage{color}
\usepackage{latexsym}
\usepackage{bbm}
\usepackage{subfigure}
\usepackage{float}
\usepackage{epsfig}
\usepackage{epsf}
\usepackage{psfrag}
\usepackage{bm}
\usepackage{amsthm}
\usepackage{eucal}
\usepackage{mathrsfs}
\usepackage{tabularx}
\usepackage{etoolbox}
\usepackage{longtable}
\usepackage{url}
\usepackage{braket}
\usepackage{epstopdf}
\usepackage{color} 

\usepackage{hyperref}
\hypersetup{
colorlinks=true,final=true,
        linkcolor=blue,
        citecolor=blue,
        filecolor=blue,
        urlcolor=blue,
}

\begin{document}

\title{Electronic structure and magnetic properties of higher-order layered nickelates: La$_{n+1}$Ni$_{n}$O$_{2n+2}$ ($n=4-6$)}
\author {Harrison LaBollita}
\affiliation{Department of Physics, Arizona State University, Tempe, AZ 85287, USA}
\author{Antia S. Botana}
\affiliation{Department of Physics, Arizona State University, Tempe, AZ 85287, USA}

\date{\today}

\begin{abstract}
The recent discovery of superconductivity in Sr-doped 
NdNiO$_2$, with a critical temperature of $10-15$ K
suggests the possibility of a new family of nickel-based high-temperature superconductors (HTS). NdNiO$_{2}$ is the $n=\infty$ member of a larger series of layered nickelates with chemical formula R$_{n+1}$Ni$_{n}$O$_{2n+2}$ (R $=$ La, Nd, Pr; $n = 2, 3, \dots, \infty$). The $n=3$ member has been experimentally and theoretically shown to be cuprate-like and a promising HTS candidate if electron doping could be achieved. The higher-order $n=4,5,$ and $6$ members of the series fall directly into the cuprate dome area of filling without the need of doping, thus making them promising materials to study, but have not been synthesized yet. Here, we perform first-principles calculations on hypothetical $n=4,5,$ and $6$ structures to study their electronic and magnetic properties and compare them with the known $n=\infty$ and $n=3$ materials. From our calculations, we find that the cuprate-like character of layered nickelates increases  from the $n=\infty$ to the $n=3$ members as the charge transfer energy and the self-doping effect due to R-$d$ bands around the Fermi level gradually decrease. 
\end{abstract}

\maketitle

\section{\label{sec:intro}Introduction}
The discovery of high-temperature superconductivity (HTS)  in cuprates in 1986 triggered an immense amount of scientific discovery \cite{highTcCuprate}. Yet, despite more than three decades of research, no consensus has emerged for the mechanism of HTS. Among different approaches to address this problem has been to look for materials with similar structures and 3$d$ electron count that are suggested as proxies for cuprate physics. In this regard, focusing on nickelates has been an obvious strategy, as Ni is next to Cu in the periodic table \cite{anisimov, pickett}.
The realization of the promise of nickelates for HTS came  with the recent report of superconductivity in Sr-doped NdNiO$_{2}$ with $T_{c} \sim 10-15 $ K \cite{Li2019, Li2020_SC_dome}. 
This observation does not only constitute the realization of the first superconducting nickel-oxide material, but also bolsters studying other layered nickelates -- NdNiO$_2$ is simply one of the members of a larger family of materials \cite{poltavets2, poltavets1}.

\begin{figure}
\centering
     \includegraphics[width=\columnwidth, trim = {0cm 0.5cm 0cm 1.25cm}, clip]{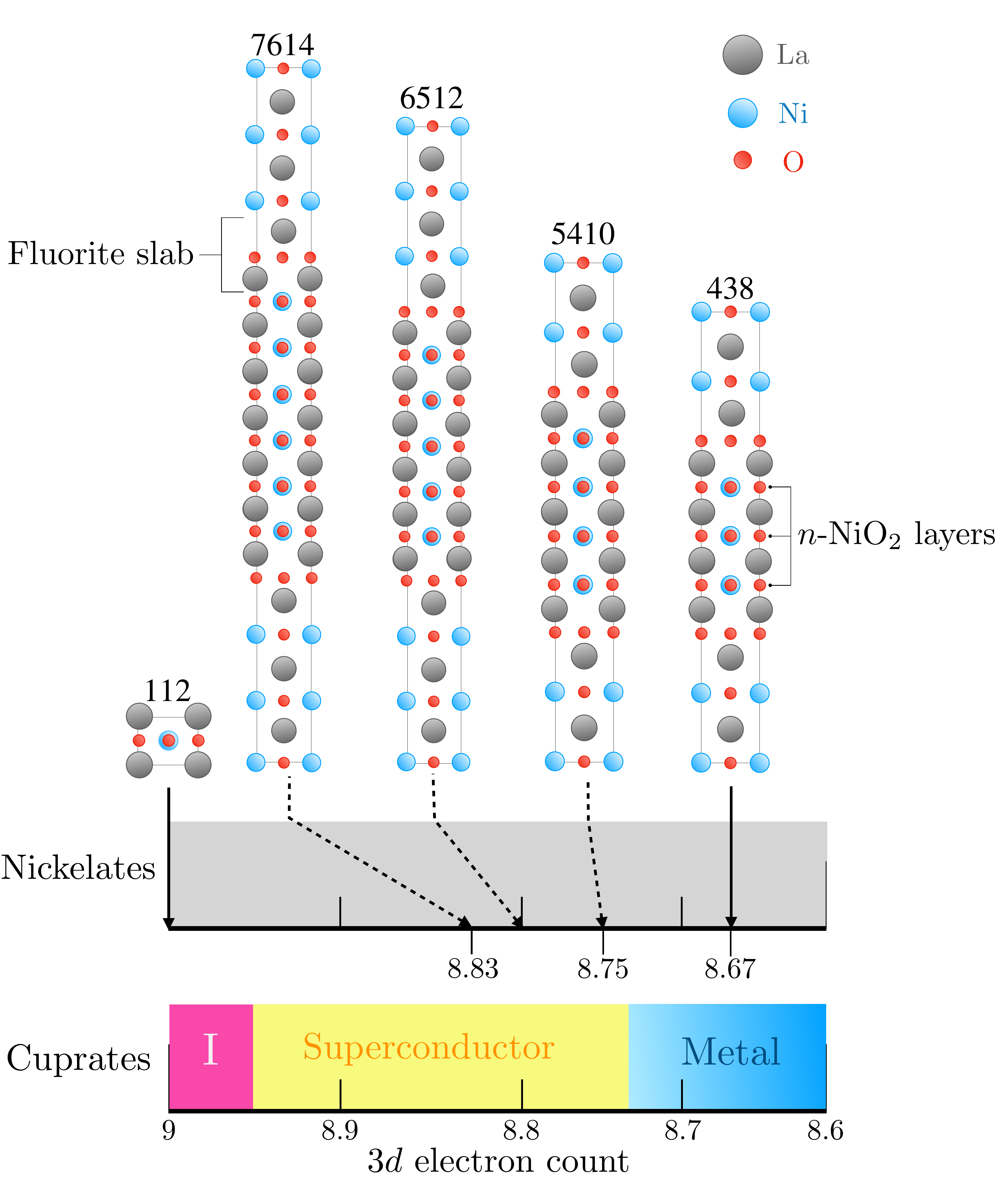}
    \caption{Schematic phase diagram of layered nickelates (top) and cuprates (bottom) presented as a function of  3$d$ electron count (I = insulator). For nickelates, only a couple line compositions are known in the range of 3$d$ electron count shown: RNiO$_2$, 112 ($n= \infty$) and R$_4$Ni$_3$O$_8$, 438 ($n=3$). Dashed lines stand for R$_5$Ni$_4$O$_{10}$, 5410 ($n=4$), R$_6$Ni$_5$O$_{12}$, 6512 ($n=5$), R$_7$Ni$_6$O$_{14}$, 7614 ($n=6$), that have not yet been synthesized yet. The structure of each of these materials is also shown. All systems contain $n$-NiO$_2$ layers and the $n=3-6$ systems also exhibit a blocking fluorite-slab.}
    \label{fig:phase}
\end{figure}

 This low-valence layered nickelate family is represented by the general formula  R$_{n+1}$Ni$_{n}$O$_{2n+2}$ (R $=$ La, Nd, Pr; $n = 2, 3,\dots, \infty$) in which each member contains $n$-NiO$_{2}$ layers, in analogy to the CuO$_2$ planes of cuprates (see Fig. \ref{fig:phase}). Layered nickelates are obtained via oxygen reduction from the corresponding parent perovskite or Ruddlesden-Popper (RP) phases \cite{poltavets1, poltavets2, GREENBLATT1997174}.
As of now, only the $n$ = 2, 3, and $\infty$ layered nickelates have been experimentally realized. NdNiO$_2$ (112) is the $n = \infty$ member of the series and realizes the hard to stabilize Ni$^{1+}$:~$d^9$ oxidation state  \cite{crespin, hayward, ikeGGda, ikeda2, Li2019}.
Superconductivity in this material is reached upon hole doping with Sr that drives the electron count into the cuprate dome area of filling (with maximum T$_c$ obtained at $d^{8.8}$ filling) \cite{Li2020_SC_dome}. This discovery has attracted a great deal of experimental \cite{Osada2020, hwang_synthesis, goodge2020, deveraux, Fu2020, Li2020, BiXiaWang2020, QiangqiangGu2020} and theoretical \cite{ES_112, arita, Liu2020, Thomale_PRB2020, Choi2020, Ryee2020, Gu2020, Leonov2020,lechermann,lechermann2, Hu2019, Sakakibara, jiang2020, werner2020, eff_ham, Zhang2020, karp2020, prx, Wang2020} attention. 
The $n=2$ member (R$_3$Ni$_2$O$_6$ -- 326)  is far from cuprates in terms of electron count \cite{poltavets2, poltavets3, greenblatt326} so we will not study it here.
The $n=3$ phases (R$_4$Ni$_3$O$_8$ -- 438), with an average Ni valence of 1.33+ ($d^{8.67}$), fall into the overdoped regime of cuprates in terms of electron count and, as such, are not superconducting even though they have been shown to be one of the closest cuprate analogs to date \cite{nat_phys, physrevmat}. Both of these nickelates (R438s and R112s) have their own respective shortcomings. In order to reach the dome area of $d$-filling, the 112 materials must be hole-doped which is known to cause structural disorder in the samples \cite{Li2019, hwang_synthesis}, the 438 materials must be electron doped which is yet to be successfully done. 

In contrast, the $n=4$, 5, and 6 members of the series  would all fall into the cuprate dome in terms of electron count without the need of doping and are hence very promising materials to pursue. Starting with the $n=4$ compound (R$_5$Ni$_4$O$_{10}$ -- 5410), it would have an average Ni valence of 1.25+ ($d^{8.75}$). The $n=5$ material (R$_6$Ni$_5$O$_{12}$ -- 6512) would have an average Ni valence of 1.2+ ($d^{8.80}$). Lastly, the $n=6$ compound (R$_7$Ni$_6$O$_{14}$ -- 7614)  would have an average Ni valence of 1.17+ ($d^{8.83}$) (see Fig. \ref{fig:phase}). Even though $n=4-6$ reduced nickelates have yet to be experimentally realized, prospects seem bright as growth of parent La-based RP $n=4-5$ phases have been recently reported \cite{Lei2017, Li2020_RP_phases}. All of these parent higher-order RP samples correspond to thin films grown via molecular beam epitaxy (MBE)-- obtaining bulk samples will likely be difficult due to thermodynamic instabilities.

\begin{figure*}
    \includegraphics[width = 2\columnwidth]{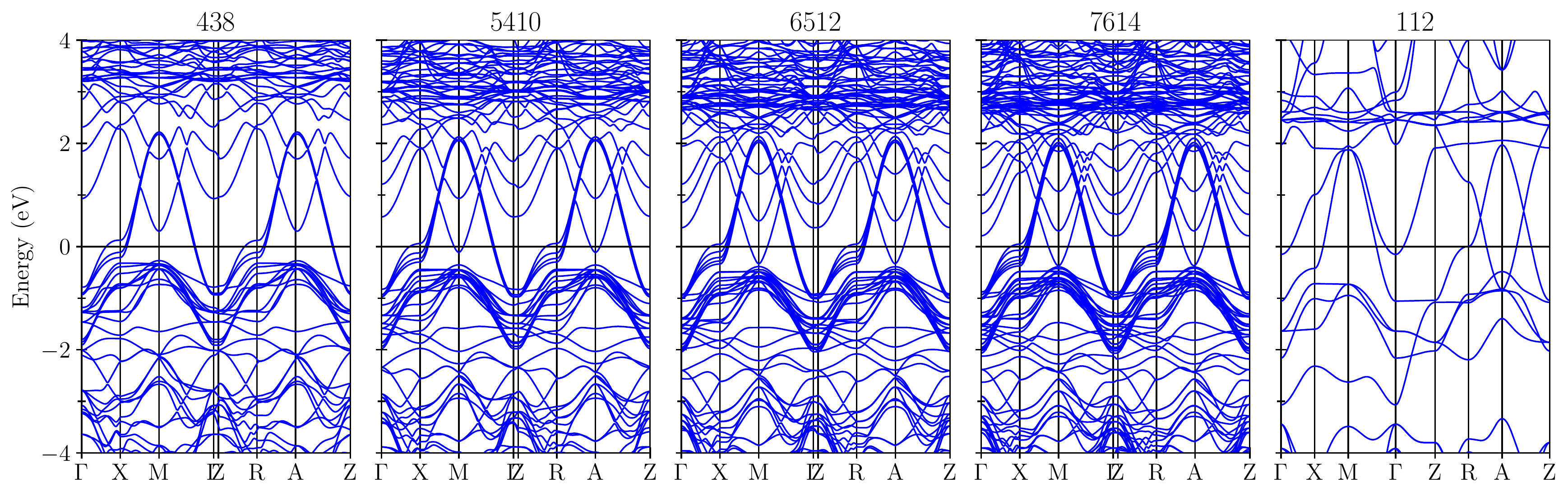}
    \includegraphics[width = 2\columnwidth]{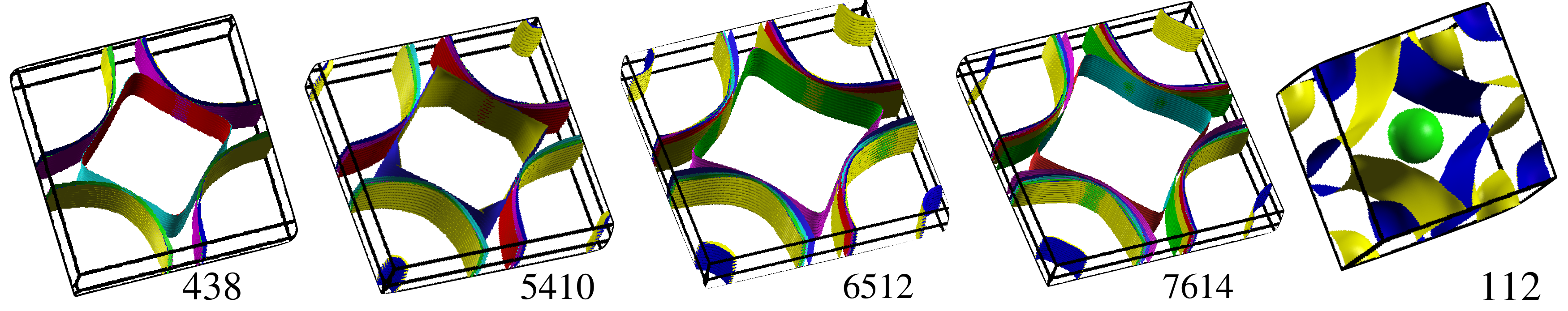}
    \caption{Top panels: Non-magnetic band structures for the La-based 438 ($n=3$), 5410 ($n=4$), 6512 ($n=5$), 7614 ($n=6$) layered nickelates (with I4/mmm space group), and for the 112 ($n=\infty$) compound (with P4/mmm space group) within GGA. Bottom panels: Corresponding Fermi surface for each compound.}
    \label{fig:nmg_bs}
\end{figure*}

Here, we use first principles calculations to describe the electronic structure of hypothetical $n=4-6$ layered nickelates (R$_{5}$Ni$_{4}$O$_{10}$, R$_{6}$Ni$_{5}$O$_{12}$, R$_{7}$Ni$_{6}$O$_{14}$ with R $=$ La), exploiting the prospect that their parent RP phases can be reduced. We then compare the electronic structure of these layered $n=4-6$ materials to the known $n=3$ and $n=\infty$ members of the series. 
We find that a low-spin state for Ni is the preferred ground state for $n=4-6$ compounds, in analogy to the 438 materials and cuprates. The R-$d$ bands (that have been reported to be relevant for the low energy physics of the 112 materials) start playing a role into the electronic structure of layered nickelates when gradually moving from the $n=3$ to the $n=6$ members. Importantly, the charge transfer energy of these compounds gradually decreases when going from $n=\infty$ to $n=3$. All in all, the cuprate-like character of low-valence layered nickelates is found to decrease with $n$.

\section{\label{sec:cm}Computational Methods}
Electronic structure calculations were performed using the all-electron, full potential code WIEN2k \cite{wien2k} based on the augmented plane wave plus local orbitals (APW + lo) basis set. The Perdew-Burke-Ernzerhof version of the generalized gradient approximation (GGA) \cite{pbe} was used for the non-magnetic calculations. 
Additionally, we perform GGA+$U$ \cite{ldau} calculations to account for the missing correlations for the Ni-$d$ states. We employ two double counting schemes in the GGA+$U$ calculations: `fully localized limit' (FLL) and `around mean field' (AMF). For both schemes we studied the evolution of the electronic structure with increasing $U$ ($U$ = 1.5 to 5.5 eV). We have chosen a non-zero $J=0.7$ eV in our calculations to properly account for the anisotropy  of the interaction \cite{pickett_lsda+u}.

We have chosen to study all layered nickelates with R $=$ La to
avoid ambiguities in the treatment of the $4f$ states that would arise from Nd or Pr. 
We note that in 112 nickelates, the La variant is, as of now, not superconducting upon hole doping (in contrast to the Pr and Nd counterparts)\cite{Li2019} even though the electronic structure of LaNiO$_2$ and NdNiO$_2$ is essentially the same (except for the $f$ bands)\cite{Sakakibara}. 
The same conclusion about the similarity of the electronic structure upon R substitution applies to the higher-order $n=4-6$ materials.
For La438, La5410 and La6512, we used a $k$-mesh of $10 \times 10 \times 10$ $k$-points in the irreducible Brillouin zone, while for the La7614 a $k$-mesh of $12 \times 12 \times 12$ $k$-points and for La112 a $k$-mesh of $9\times9\times11$ $k$-points was required for convergence. We used $R_{\text{MT}}K_{\text{max}} = 7.0$ and muffin-tin radii 2.50, 1.99, and 1.72 \AA\  
for La, Ni, and O, respectively, for all of our calculations. 

\section{\label{sec:struct}Structural Properties}
We construct the structure of hypothetical La-based $n=4-6$ nickelates using the structure of the La$_4$Ni$_3$O$_8$ material as a reference. The structure of 438 materials is tetragonal (with an I4/mmm space group) and contains three NiO$_2$ planes separated by a single layer of R ions, similar to the 112 materials. However, in contrast to infinite-layer nickelates (with P4/mmm space group), in the 438 materials, the NiO$_2$ planes are separated along the $c$-axis by a fluorite-like RO blocking slab (see Fig. \ref{fig:phase}) that makes the inter-trilayer coupling very weak (this fluorite slab is absent in the 112 phase). We safely assume that higher-order nickelates  will have an analog structure to the 438 materials once synthesized i.e., tetragonal with an I4/mmm space group, $n$-NiO$_{2}$ planes along $c$, and fluorite slabs formed by the rare-earth and oxygen ions as spacing layers. With these considerations, we have all of the ingredients to construct the structure for each of the reduced La-based $n=4-6$ higher-order phases. We then optimize the lattice parameters and internal coordinates for each phase. Our optimizations were done using a ferromagnetic configuration within GGA, the ground state for these systems (see below).

The optimized lattice parameters for each higher-order phase are shown in Table \ref{tab:latt_params}, those for La438 and La112 compounds are also shown as a reference. The in-plane lattice parameters are almost identical for all compounds whereas the out of plane lattice parameter obviously increases with the number of layers. Relevant bond lengths are consistent for all compounds: d$_{\text{Ni-O}}= 1.97$ -- 1.99 \AA, d$_{\text{La-Ni}}= 3.15$ -- 3.26 \AA, and d$_{\text{La-O}}= 2.51$ -- 2.59 \AA. 
All in all, from our structural relaxations, we find the obtained bond lengths agree well with the experimentally derived values for the La438 and La112 compounds which serves as a benchmark for our relaxations in the $n=4-6$ structures \cite{poltavets1, ZhangPNAS}.

\begin{table}
    \centering
    \begin{tabular*}{\columnwidth}{l@{\extracolsep{\fill}}lccc}
    \hline
    \hline 
     $n$ & Material &  $a$ & $b$ & $c$ \\
     \hline
     3 & La438  & 3.97 & 3.97 & 26.1\\
     4 & La5410 & 3.97 & 3.97 & 32.9\\
     5 & La6512 & 3.96 & 3.96 & 39.9\\
     6 & La7614 & 3.97 & 3.97 & 46.1\\
     $ \infty$ & La112 & 3.95 & 3.95 & 3.37\\
     \hline 
     \hline
    \end{tabular*}
    \caption{Calculated lattice parameters for higher-order nickelates (5410, 6512, 7614), as well as experimental 438 \cite{poltavets1} and 112 \cite{hayward} lattice parameters. All values are given in \AA.}
    \label{tab:latt_params}
\end{table}

\section{\label{sec:nmgc}Non-Magnetic Calculations}

Fig. \ref{fig:nmg_bs} shows the non-magnetic band structures and corresponding Fermi surfaces of the La-based $n=4-6$ nickelates after the structural relaxations (the band structures and Fermi surfaces for La438 and La112 using the experimental structural data are shown as a reference). In cuprates, a single band of $d_{x^2-y^2}$ character hybridized with O-$p$ states is active in the vicinity of the Fermi level.  For the La112 compound (at $d^9$ filling), in addition to the $d_{x^{2} - y^{2}}$ band, there are also La-$5d$ bands crossing the Fermi level at both A (with dominant $d_{xy}$ character) and $\Gamma$ (with $d_{z^2}$ character). As such, the corresponding Fermi surface of the 112 material contains not only a cylinder with Ni-$d_{x^2-y^2}$ holes, but also an electron-like spherical pocket in the center ($\Gamma$, with La-$d_{z^2}$ character) and another electron-like sphere at each corner (A, with La-$d_{xy}$ character). This has been described in earlier literature and quite some attention has been paid to the role of R-$d$ bands, as they make the electronic structure and fermiology non-cuprate-like in R112 materials \cite{prx, pickett, arita, doped_MI, ES_112}.

The evolution of the electronic structure of La-based layered nickelates from $n=\infty$ to $n=3$ shows one $d_{x^2-y^2}$ band per Ni crossing the Fermi level for each material, as expected (see Fig. \ref{fig:nmg_bs}), with a bandwidth that does not significantly change with $n$. In the multi-layered systems ($n=3-6$), a splitting between the Ni-$d_{x^2-y^2}$ bands is observed at X as a consequence of interlayer hopping, similar to that in multi-layer cuprates \cite{Sakakibara2014}. The most important difference as $n$ increases is the gradual involvement of La-$d$ bands. In the 438 ($n=3$) compound  the situation is very similar to that in the cuprates: there is a single $d_{x^{2} - y^{2}}$ band per Ni crossing the Fermi level, with no La-$d$ involvement, as reported before \cite{nat_phys}. In La5410 La-$d$ bands start crossing the Fermi level giving rise to electron pockets at A and M, and they gradually shift down to lower energies for La6512 and La7614. Unlike in the 112 material, there is no La-$d$ band crossing at $\Gamma$.

These differences in electronic structure can be easily appreciated in the corresponding Fermi surfaces also shown in Fig. \ref{fig:nmg_bs}. In La438 ($n=3$) the Fermi surface consists of three Ni-$d_{x^2-y^2}$-derived pockets only: the
bonding, non-bonding, and antibonding superpositions of
the three layers --  the outer
hole-like pockets centered around the zone corner correspond to the lower-lying bands, while the inner pocket (nearly square-like around $\Gamma$) arises from the antibonding (higher-lying) band. These Ni-$d_{x^2-y^2}$ derived pockets are kept in the $n=4-6$ materials -- with extra hole pockets centered around the zone corner as the number of Ni-$d_{x^2-y^2}$ bands crossing the Fermi level increases with $n$. The important difference arises from the La-$d$ bands crossing the Fermi level at A and M in the $n=4-6$ systems mentioned above-- the La-$d$ electron-like spherical pocket at A in the 112 compound becomes cylindrical in the $n=4-6$ materials. The size of this cylindrical pocket increases with $n$. The small spherical pocket at $\Gamma$ in the 112 compound is absent in the $n=4-6$ materials (as mentioned above, unlike in the 112 material, there is no La-$d$ band crossing at $\Gamma$ in the $n=4-6$ systems). One should note that the structure of the $n=3-6$ nickelates is different from that of the infinite-layer material as the former have a fluorite blocking layer that cuts the $c$-axis dispersion, and causes the changes in the electronic structure and fermiology we have described.  
Importantly, the Fermi surfaces of the $n=3-6$ nickelates are very similar to those of multi-layered cuprates (see for example Ref. \onlinecite{Sakakibara2014}), except for the additional contributions from the R-$d$ bands in the $n=4-6$ compounds.


\begin{figure}
    \centering
    \includegraphics[width = \columnwidth]{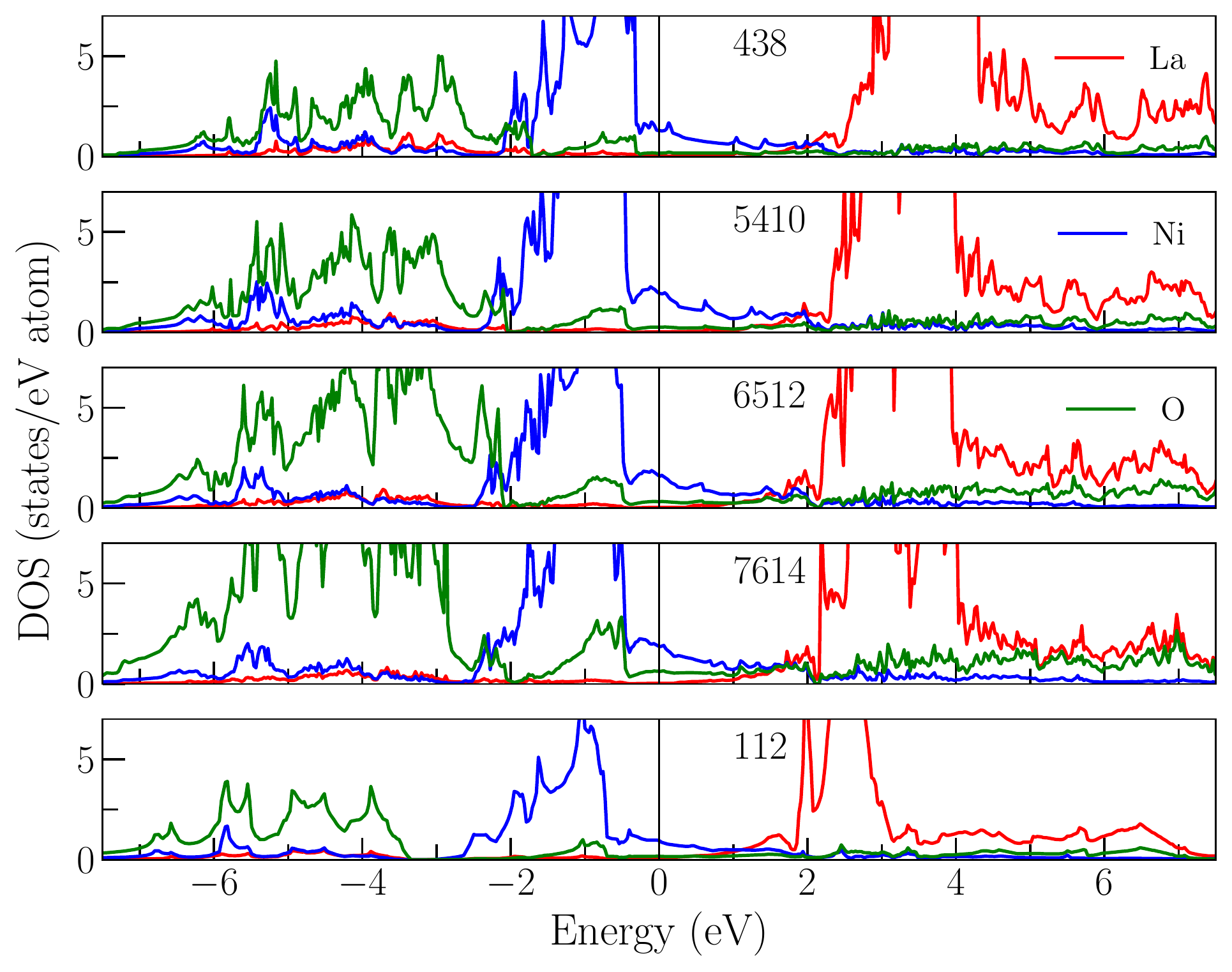}
    \caption{Comparison of the total non-magnetic atom-resolved density of states for the La, Ni, and O ions in La-based 438 ($n=3$), 5410 ($n=4$), 6512 ($n=5$), 7614 ($n=6$), and 112 ($n=\infty$) layered nickelates within GGA.}
    \label{fig:nmg_dos}
\end{figure}

\begin{figure}
    \centering
    \includegraphics[width = \columnwidth, trim = {0.125cm 0.125cm 1cm 1cm}, clip]{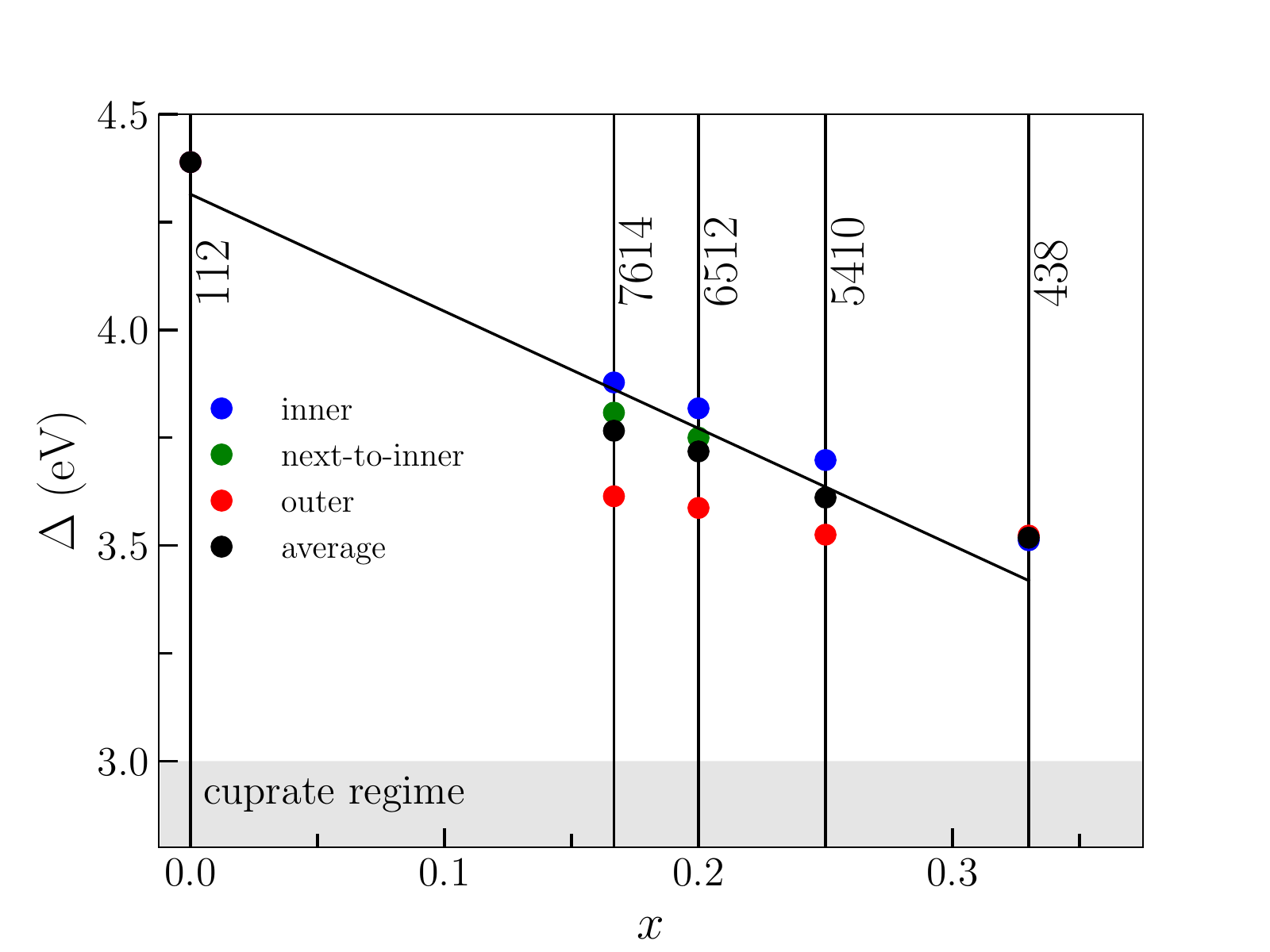}
    \caption{Charge transfer energies (average and layer-dependent) plotted versus doping level ($x= 1/n$) for each La-based layered nickelate ($x= 0$ for La112, $x= 0.17$ for La7614, $x= 0.2$ for La6512, $x= 0.25$ for La5410, and $x= 0.33$ for La438). }
    \label{fig:ct_v_hole}
\end{figure}

To gain a complete description of the non-magnetic electronic structure, Fig. \ref{fig:nmg_dos} shows the atom-resolved densities of states (DOS) for the La, Ni, and O ions within GGA for $n=\infty$ and $n=3-6$ layered nickelates. The density of states plots confirm the above description and clearly show how the La-$5d$ states shift down in energy as $n$ increases. The centroid of the Ni-$d$ states does not significantly change across the series. In contrast, the O-$p$ centroid shifts to lower energies (away from the Fermi level) when going from the La438 compound to the La112 compound (by $\sim$ 1 eV across the series). As a consequence, the degree of $p-d$ hybridization increases gradually from La112 to La438 \cite{dean}. In connection to this, studies of 112 materials have highlighted their much larger charge transfer energy ($\Delta = \varepsilon_{d} - \varepsilon_{p}$) with respect to cuprates ($\Delta_{112}$ $\sim$ 4 eV whereas prototypical cuprate values are $\sim$ 2 eV \cite{weber}). This is a relevant parameter in cuprates, as many theories of HTS in cuprates are based on the large degree of $p-d$ hybridization, that ultimately allows for Zhang-Rice singlet formation \cite{zhangRice}. Importantly, a decreasing charge transfer energy across cuprates has been shown to result in higher $T_{c}$ values \cite{weber}. 
To obtain a quantitative estimate for each nickelate, explicit $\Delta$ values are obtained following Ref. \onlinecite{gw} using band centroids (for Ni-$d_{x^2-y^2}$ and O-$p\sigma$) calculated as:
\begin{equation}
    \label{eqn:centroid}
    E_{i} = \frac{ \int E g_{i}(E) \, dE}{ \int g_{i}(E) \, dE},
\end{equation}
where $g_{i}(E)$ is the partial density of states for orbital $i$. An ambiguity can inevitably be introduced in determining the integration range -- we set it to cover the bonding-antibonding band complex for Ni-$d_{x^2-y^2}$  states, and the DOS range for O-$p\sigma$, as the values obtained for the centroids in 112 and 438 materials using such integration limits are similar to  the on-site energies of the corresponding maximally localized Wannier functions reported in the literature \cite{prx, nica2020}.

\footnotetext[1]{We attempted to obtain estimates of the charge transfer energies for the $n=3-6$, and $\infty$ compounds from the on-site energies of maximally localized Wannier functions (Ni-$d_{x^2-y^2}$ and O-$p\sigma$) derived using Wannier90 \cite{wannier90, wien2wannier}. We obtain well-localized Wannier functions (with spreads $\sim 1$\AA$^{2}$) for the $n=3,4$ and $\infty$ materials using an energy window of -8 eV to 3 eV with the initial orbital projections being La-$d_{xy}$, La-$d_{z^{2}}$, all Ni-$d$, and O-$p$ orbitals. For the $n=5$ and 6 materials, we could not obtain well-localized Wannier functions. Therefore, we present the charge transfer energies obtained from the density of states-derived centroids for all materials for consistency.}

The derived average centroid energies (with respect to the Fermi level) for Ni-$d_{x^{2}-y^{2}}$ and O-$p_{\sigma}$ orbitals, as well as estimated charge transfer energies are summarized in Table \ref{tab:ct_table}. The average charge transfer energy for each material increases with $n$ with an overall increase of $\sim$ 1 eV as could be inferred from the DOSs. We note that the charge transfer energies calculated from the density of states-derived centroids agree well with those obtained using the on-site energies from maximally localized Wannier functions for the $n=3, 4$, and $\infty$ materials (see Table \ref{tab:ct_table} and Ref. \onlinecite{Note1}).

As mentioned above, many theories of HTS in cuprates are based on their small charge transfer energy. If this is a relevant parameter in layered nickelates as well, $n=3-6$ phases should then be promising materials to explore.

\begin{table}
\begin{tabular*}{\columnwidth}{c@{\extracolsep{\fill}}cccc}
\hline 
\hline
$n$ & E$_{\text{O-}p\sigma}$ (eV) & E$_{\text{Ni-}d_{x^2-y^2}}$ (eV) & $\Delta_{\text{DOS}}$ (eV) & $\Delta_{\text{Wannier}}$ (eV)\\
\hline
3        & -4.53 & -1.02  &  3.52 & 3.52 Ref. \onlinecite{nica2020}\\
4        & -4.65 & -1.04  &  3.61 & 3.80 Ref. \onlinecite{Note1}\\
5        & -4.79 & -1.07  &  3.72 & -- \\
6        & -4.85 & -1.05  &  3.80 & -- \\ 
$\infty$ & -5.51 & -1.12  &  4.39 & 4.40 Ref. \onlinecite{prx}\\
\hline
\hline
\end{tabular*}
\caption{Estimates of the on-site energies of the Ni-$d_{x^{2}-y^{2}}$ and O-$p_{\sigma}$ orbitals from the DOSs and the estimated charge transfer energy. Additionally, we include estimates of the charge transfer energy obtained via wannierizations for the $n=3$, 4 and $\infty$ compounds using Wannier90 \cite{wannier90} and wien2wannier\cite{wien2wannier}.} 
\label{tab:ct_table}
\end{table}

Fig. \ref{fig:ct_v_hole} shows the charge transfer energy (average and layer-dependent) versus doping $x=1/n$. This plot reflects that a linear fit to the average charge transfer energy is good, even though it shows some slight deviations, indicating some sensitivities in our estimates. The 112 has effectively a single layer, the 438 and 5410 have inner and outer layers, and the 6512 and 7614 have inner, next-to-inner, and outer layers. Interestingly, we find the charge transfer energy is layer-dependent in a systematic manner within each material: larger on the inner NiO$_{2}$ layers, decreasing gradually as we move outwards.

\section{\label{sec:mgc}Magnetic Calculations}

Using spin-polarized calculations, we have performed
a full study of the stability of different spin states for
different magnetic configurations in all layered nickelates constructing $\sqrt{2}\times\sqrt{2}$ supercells (with Cmmm space group). 

Concerning possible Ni spin states in these materials, one should start by considering that, for $n=3-6$ nickelates, the Ni ions sit in a square-planar environment with an average $d$-filling that can be described as $d^{8 + \delta}$ (with $\delta$ increasing with $n$). This is also effectively the picture in the $n=\infty$ compound (even though a simple ionic count gives a Ni$^{1+}$: $d^9$) due to the self-doping effect from R-$d$ bands. Importantly, there are two possible spin states for such a Ni: low-spin (LS) and high-spin (HS). These two possibilities arise from the competition between the crystal field splitting ($\Delta_{\text{CF}}$) in the e$_{g}$ states and Hund's rule coupling ($J_{\text{H}}$). The LS state (with $\Delta_{\text{CF}}$ larger than $J_{\text{H}}$) corresponds to a $d$-filling (t$_{2g}$)$^{6}$($d_{z^{2}}$)$^{2}$($d_{x^{2} - y^{2}}$)$^{\delta}$ with $S = \delta/2$ and moment of $\delta$ per nickel. The HS configuration (with $J_{\text{H}}$ larger than $\Delta_{\text{CF}}$) corresponds to a $d$-filling (t$_{2g}$)$^{6}$($d_{z^{2}, \uparrow}$)$^{1}$($d_{x^{2} - y^{2},\uparrow}$)$^1$ ($d_{z^2,\downarrow}$)$^{\delta}$ with $S = (2 - \delta)/2$ and a moment of $2-\delta$ per nickel. Understanding the preferred spin state is crucial to understanding the physics of these materials. Specifically, a HS spin state is non-cuprate-like as it involves Ni-$d_{z^2}$ states around the Fermi level, while a LS spin state is cuprate-like with the involvement of Ni-$d_{x^{2} - y^{2}}$ states only.


\begin{figure}
    \centering
    \includegraphics[width = \columnwidth, trim = {0.25cm 0.25cm 0.25cm 0.15cm}, clip]{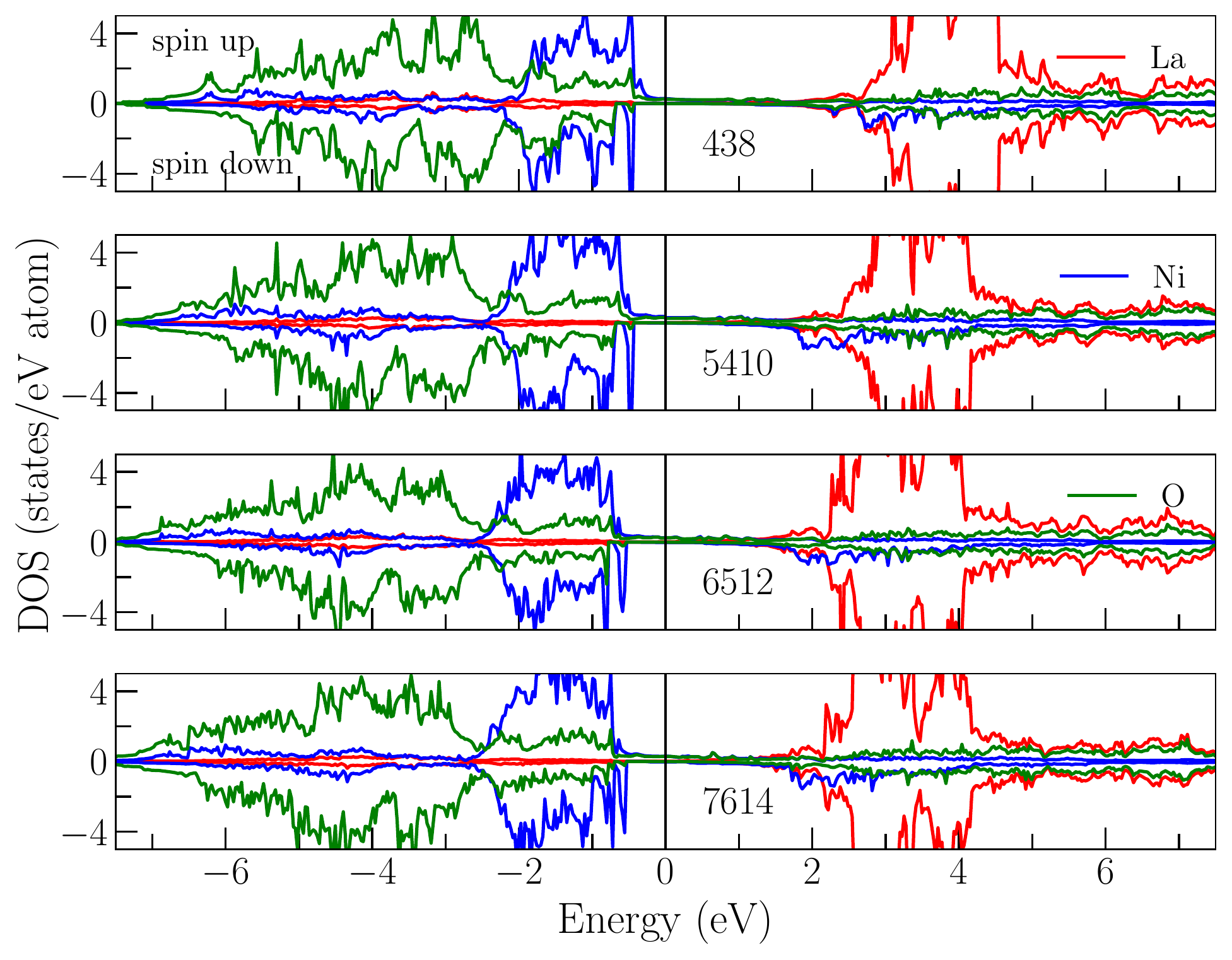}
    \caption{Comparison of the atom-resolved densities of states for the La, Ni, and O ions in La-based 438 ($n=3$), 5410 ($n=4$), 6512 ($n=5$), and 7614 ($n=6$) layered nickelates in their LS-FM metallic ground state within GGA+$U$ (AMF) for $U = 5.5$ eV.}
    \label{fig:fm_dos_ggau} 
\end{figure}

In order to find the ground state for each compound, we first compare the energies of the antiferromagnetic (AFM), ferromagnetic (FM), and non-magnetic states at the GGA  level. Within GGA, the energy of the non-magnetic state for all compounds is higher compared to both the AFM and FM configurations by $10 - 60$ meV/Ni. The energy differences and Ni magnetic moments (LS for all materials) from GGA calculations on $n=3-6$ and $\infty$ nickelates are shown in Appendix \ref{sec:mm} Table \ref{tab:GGA}.

Given that the non-magnetic state is higher in energy than the FM or AFM in all materials, we investigate the relative stability between the latter states at the GGA+$U$ level with the two double counting schemes mentioned above (AMF and FLL). We note that AMF and FLL are known to stabilize different spin states:  FLL tends to stabilize HS states whereas AMF tends to penalize magnetic energies and hence has a tendency to stabilize LS states instead \cite{pickett_lsda+u}. The energy differences ($E_{\text{AFM}} - E_{\text{FM}}$) and Ni magnetic moments for each compound from all of our GGA+$U$ calculations have been tabulated in Appendix \ref{sec:mm} Tables \ref{tab:GGAU_FLL} (FLL) and \ref{tab:GGAU_AMF}  (AMF). In a high-spin configuration for the Ni $d^{8+\delta}$ ions (with half-filled $d_{x^2-y^2}$ bands) an AFM checkerboard configuration is preferred. In contrast, the low-spin Ni $d^{8+\delta}$ ions (with $d_{x^2-y^2}$ bands away from half-filling) prefer a FM order.
Within both the AMF and FLL schemes, our calculations show a LS-FM ground state is preferred for the $n=4-6$ compounds. The stability of this state has been confirmed using fixed spin moment calculations. This makes these higher-order phases similar to the R438 ($n=3$) material for which a LS-FM ground state is also obtained  from first-principles calculations that has been shown to be consistent with polarized x-ray absorption experiments \cite{nat_phys}. 
The LS state is metallic, with all Ni moments remaining below 1$\mu_{B}$, and the involvement of Ni bands of $d_{x^2-y^2}$ character-only around the Fermi level (see below). 
In contrast to $n=3-6$ materials, for La112 ($n=\infty$) we find instead an AFM ground state with a LS-to-HS transition upon increasing $U$, in agreement with previous literature\cite{prx, Kapeghian2020, pickett} 
This AFM state in La112 remains metallic up to the highest $U$ value used, and is fundamentally different due to the involvement of a flat Ni-$d_{z^2}$ band (in addition to the $d_{x^2-y^2}$) at the Fermi level \cite{pickett3}. 

For metallic R438, the AMF scheme has been shown to give a description of the electronic structure that is in much better agreement with experiments\cite{nat_phys}. Given their similarities, we expect the same for the $n=4-6$ materials so we adopt this scheme to describe 
the electronic structure of their LS metallic ground state. Fig. \ref{fig:fm_dos_ggau} shows the  atom-resolved densities of states for the La, Ni, and O ions for $n=4-6$ nickelates in their ground state (for a standard $U$ for Ni of 5.5 eV) \cite{botana2016, pardo2010, nat_phys} -- for the corresponding band structures see Fig. \ref{fig:fm_afm_bs} in  Appendix \ref{sec:fm_bs}. The most important feature of the solution we have obtained for $n=4-6$ materials is the LS state nature of the Ni ions with the sole involvement of one Ni-$d_{x^2-y^2}$ band per Ni around the Fermi level. As mentioned above, this contrasts with the nature of the high-spin AFM ground state obtained in the $n=\infty$ compounds \cite{pickett3, Kapeghian2020}, that is characterized by a multi-orbital nature with Ni-$d_{z^2}$ states pinned at the Fermi level playing a crucial role. In all cases ($n=4-6$ and $\infty$), R-$d$ bands still cross the Fermi level but the LS ground state of the $n=4-6$ materials is more cuprate-like in nature as only Ni-$d_{x^2-y^2}$ states play a role in the low energy physics. Overall, the trends with $n$ of the electronic structure in the LS state  are similar to those of the non-magnetic state: i) there is a shift of the La-$5d$ states to lower energies as $n$ increases across the series (more evident in  Fig. \ref{fig:fm_afm_bs}), ii) the centroid of the Ni-$d$ states remains nearly fixed throughout the series, iii) the O-$p$ states shift to lower energies with increasing $n$ (by $\sim$ 1 eV across the series). As discussed earlier, these effects collectively decrease the degree of $p-d$ hybridization as $n$ increases.  

Concerning the LS-FM nature of the ground state we have described for $n=4-6$ materials, it is important to note that antiferromagnetic states are observed in cuprates near $d^{9}$ filling (indeed, we find an AFM ground state for the $n=\infty$ material). However, when hole doped away from $d^9$, one finds broken symmetry phases in both the cuprates and layered nickelates\cite{Tranquada1996, poirot, Yi1998, wochner}. While we have obtained a LS-FM ground state for $n=4-6$ nickelates, earlier work on R438s ($n=3$) found that this LS state does in fact compete in energy with a charge and spin-striped phase  \cite{438_ss, nat_phys}. With the higher-order $n=4-6$ materials exhibiting a similar electronic structure to the R438s ($n=3$), we anticipate such broken symmetry phases or even complex magnetic behavior \cite{pr438_spinglass} may be present in these materials as well.
Given that charge/spin-stripe phases would require large supercells, their analysis is beyond the scope of this work and we leave this for future calculations and experimental observations. We point out that a nesting-driven charge/spin-density wave transition has actually been previously commented on by Poltavets et al.\cite{GreenblattPRL} for the $n=3$ nickelate based on its fermiology. Given that the Fermi surfaces of the $n=3$ and $n=4-6$ materials are indeed similar (see Fig. \ref{fig:nmg_bs}), this can be a likely possibility in these latter phases as well.

Finally, even though the AFM state is not the ground state in $n=4-6$ materials, we briefly describe some interesting aspects of it here. First, a LS-to-HS transition with $U$ is obtained within both AMF and FLL schemes (akin to that in La112). However, in contrast to La112, at high enough $U$ values this AFM state in $n=3-6$ nickelates is insulating, specifically, at $U$ $\sim$ 2.7 eV for La438 and $U \sim 5.5$ eV for La5410, La6512, and La7614. We attribute the possibility to open up a gap in the AFM state in $n=3-6$ materials (in stark contrast to the 112s) to the existence of the blocking fluorite-slab that cuts the $d_{z^2}$ off-plane hopping. Interestingly, in the AFM state for $n=4-6$ nickelates there is a tendency for the moments to disproportionate between inner and outer layers, with disproportionation being more pronounced in the La5410 and La7614 cases, that lack a mirror plane. The higher Ni moments are found at the inner layers and gradually decrease as moving outwards. This effect is present in both double counting schemes, the FLL scheme only exacerbates it (see Appendix \ref{sec:mm} Table \ref{tab:GGAU_FLL}). This disproportionation effect has been seen before in multi-layered cuprates, where there is a charge imbalance between the inner versus outer planes \cite{charge_imb_1, charge_imb_2, charge_imb_3, charge_imb_4, charge_imb_5}. 



\section{\label{sec:summ} Conclusions}

We have used first-principles calculations to study the electronic and magnetic properties of hypothetical La-based $n=4-6$ layered nickelates and compared them with those of the known $n=3$ and $n=\infty$ materials. Our calculations show interpolation between cuprate-like and 112-like behaviour as $n$ increases in these low-valence layered nickelates. In particular, we find that a low-spin state for Ni is the preferred ground state of the $n=4-6$ compounds analogous to the 438 materials and cuprates. The  self-doping effect due to R-$d$ bands and the charge transfer energy of these materials gradually increase with $n$. As many theories of HTS in cuprates are based on their large degree of $p-d$ hybridization (small charge-transfer energy) we argue that $n=3-6$ layered nickelates are then very promising materials to study in this context.

\section*{Acknowledgements}
 We thank  M. R. Norman for fruitful discussions. We acknowledge the support from NSF-DMR 2045826 and from the ASU Research Computing Center for HPC resources.

\bibliography{bib}

\appendix

\section{Energies and Magnetic Moments}
\label{sec:mm}
The energy differences between different magnetic configurations within GGA and GGA$+U$ calculations are described in this Appendix.  Table \ref{tab:GGA} shows the results from our GGA calculations. We include the Ni magnetic moments for both AFM and FM spin configurations, as well as energy differences between the non-magnetic (NM) and AFM/FM states. In all cases, the FM and/or AFM states are favored over the NM state across the series. The results from our GGA$+U$ calculations are summarized in Tables \ref{tab:GGAU_FLL} and  \ref{tab:GGAU_AMF}. In Table \ref{tab:GGAU_FLL}, we provide the energy differences between the AFM state and the FM state with GGA+$U$ for a range of $U$ values with the FLL double counting scheme. A qualitative description of the ground state (GS) of each compound, and the Ni magnetic moments for the AFM and FM configurations in all layers are also provided. The same information is provided in Table \ref{tab:GGAU_AMF}  for the AMF double counting scheme.

\begingroup
\begin{table*}
\centering
\begin{tabular*}{2\columnwidth}{c@{\extracolsep{\fill}}cclccc}
\hline 
\hline 
Material & $E_{\text{NM}} - E_{\text{AFM}}$ (meV/Ni) & $E_{\text{NM}} - E_{\text{FM}}$ (meV/Ni) &  NiO$_{2}$ layer & AFM MM  & FM MM \\
\hline 
La438 & 6.36 & 23.32 & inner & 0.30/-0.30 & 0.51\\
      &      &       & outer & 0.27/-0.27  & 0.50\\ 
\hline
La5410 & 25.52 & 19.52  & inner & 0.44/-0.44 & 0.51\\
       &       &              & outer & 0.34/-0.33 & 0.49\\
\hline       
La6512 & 40.51 & 13.92 &  inner & 0.54/-0.54 & 0.44\\ 
       &       &         & next-to-inner & 0.48/-0.48 & 0.47\\
       &       &       &     outer & 0.37/-0.37 & 0.49 \\
\hline 
La7614 & 66.25 & 16.79 &  inner & 0.59/-0.59 & 0.46\\
       &       &       &  next-to-inner & 0.51/-0.51 & 0.52 \\
       &       &       &         outer & 0.39/-0.39 & 0.55\\
\hline
La112 & 66.81 & -2.04 &  & 0.68/-0.68 & 0.31\\
      &       &       &  &            &     \\
\hline 
\hline
\end{tabular*}
\caption{Energy differences (in meV/Ni) between non-magnetic (NM) and antiferromagnetic (AFM)/ferromagnetic (FM) spin configurations within GGA  for $n= 3-6$ and $\infty$ materials. Additionally, the magnetic moments (MM) for both the antiferromagnetic and ferromagnetic spin configurations are shown for Ni atoms in each distinct layer. Magnetic moments are in $\mu_{B}$.}
\label{tab:GGA}
\end{table*}
\endgroup

\begingroup
\begin{table*}
\centering
\caption{{\bf GGA+$U$ (FLL)} --  Energy differences (in meV/Ni) between a checkerboard antiferromagnetic (AFM) and ferromagnetic (FM) state for La-based layered nickelates with $n=3$, 4, 5, 6, and $\infty$ within GGA+$U$ as a function of $U$ using the FLL double counting scheme. A positive energy difference indicates the FM state is energetically favored, while a negative difference indicates the AFM state is favored. A qualitative description of the ground state (GS) of the system at each value of $U$ is provided. Finally, the Ni magnetic moments for both AFM and FM spin configurations are given for each Ni in each layer. (-) indicates a calculation that could not be converged.}
\label{tab:GGAU_FLL}
\begin{tabular*}{2\columnwidth}{c@{\extracolsep{\fill}}clccc}
    \hline
    \hline
    $U$ (eV)  &  $E_{\text{AFM}}-E_{\text{FM}}$ (meV/Ni) & GS  & NiO$_{2}$ Layer & AFM Moments ($\mu_{B}$) & FM Moments ($\mu_{B}$) \\
    \hline
    \multicolumn{6}{c}{{ \bf La438}} \\
\hline
    1.5 & -13.8 & HS-AFM  & inner &  1.1/-1.1 & 0.68 \\
        &     & metal & outer &  0.92/-0.92 & 0.67 \\
\hline
    2.7  & -107.6 & HS-AFM        & inner & 1.3/-1.3 & 0.72  \\
         &        & insulator     & outer & 1.1/-1.1 & 0.71  \\
\hline
    4.0 & -216.9 & HS-AFM   & inner & 1.4/-1.4 & 1.3 \\
        &      & insulator  & outer & 1.2/-1.2 & 0.87\\
\hline
    5.5 & -281.7 & HS-AFM & inner & 1.4/-1.4 & 1.5\\
        &   & insulator    & outer & 1.3/-1.3 & 1.3\\
\hline 
    \multicolumn{6}{c}{{ \bf La5410} }\\
    \hline
    1.5 & 91.5 & LS-FM & inner & 0.97/-0.97 & 0.76 \\
        &      & metal & outer &  0.54/-0.56 & 0.73  \\
    \hline
    2.7 & 114.8 & LS-FM & inner & 1.2/-1.2 & 0.80 \\
        &      & metal & outer & 0.71/-0.71 & 0.77 \\
    \hline 
    4.0 & 59.9 & LS-FM & inner & 1.3/-1.3 & 0.84 \\
        &     &  metal     & outer & 0.78/-0.78 & 0.81\\
    \hline 
    5.5 & 79.2 & HS-FM & inner & 1.4/-1.4 & 1.4\\
        &      & metal & outer & 0.93/-0.93 & 1.3\\
    \hline 
    \multicolumn{6}{c}{{ \bf La6512} }\\
    \hline
    1.5 & 48.1 & LS-FM & inner & 0.69/-0.69 & 0.80  \\
        &      & metal & next-to-inner & 0.65/-0.65 & 0.78 \\
        &      &       & outer & 0.55/-0.54 & 0.75 \\
    \hline
    2.7 & 134.7 & LS-FM & inner & 0.90/-0.90 & 0.85\\
        &       & metal & next-to-inner & 0.82/-0.82 & 0.83\\
        &       &       & outer         & 0.69/-0.70 & 0.79\\
    \hline 
    4.0 & 79.6 & LS-FM & inner & 1.2/-1.2 & 0.89\\
        &       & metal & next-to-inner & 0.61/-0.61 & 0.88\\
        &       &       & outer         & 1.2/-1.2 & 0.84\\
    \hline
    5.5 & 19.5 & LS-FM & inner & 1.3/-1.3 & 0.94\\
        &      & metal & next-to-inner & 0.57/-0.57 & 0.92\\
        &       &       & outer         & 1.3/-1.3 & 0.87\\
    \hline
    \multicolumn{6}{c}{{ \bf La7614} }\\
    \hline
    1.5 & 13.4 & LS-FM & inner & 0.89/-0.89 & 0.83\\ 
        &      & metal & next-to-inner & 0.64/-0.64 & 0.80 \\
        &       &      & outer & 0.63/-0.63 & 0.76\\
    \hline
    2.7 & 56.3 & LS-FM & inner & 1.1/-1.1 & 0.87\\
        &       & metal & next-to-inner & 0.64/-0.64 & 0.85\\
        &       &       & outer         & 1.1/-1.1 & 0.81 \\
    \hline
    4.0 & 32.8 & LS-FM & inner & 1.2/-1.2 & 0.92\\
        &       & metal & next-to-inner & 0.67/-0.67 & 0.90\\
        &       &       & outer         & 1.2/-1.2 & 0.85 \\
    \hline
    5.5 &    &   & inner & 1.3/-1.3 & - \\
        &     &    & next-to-inner & 0.66/-0.66 & - \\
        &     &    & outer         & 1.3/-1.3  & -\\
    \hline
  \multicolumn{6}{c}{{ \bf La112}}\\
  \hline
    1.5 & -143.8 & LS-AFM &  & 0.83/-0.83 & 0.87 \\
        &        & metal  &    &            &  \\
    \hline
    2.7 & -138.5 & HS-AFM &  & 1.04/-1.04 & 0.93\\
        &       & metal  &    &            &      \\
    \hline
    4.0 & -146.4 & HS-AFM &  & 1.1/-1.1 & 0.99 \\
        &       & metal  &    &            &   \\
    \hline
    5.5 & -135.4 & HS-AFM &   & 1.2/-1.2 & 1.1 \\
        &       &   metal     &    &    & \\
    \hline  
    \hline
\end{tabular*}

\end{table*}
\endgroup

\begingroup
\begin{table*}
\caption{{\bf GGA+$U$ (AMF)} --  Energy differences (in meV/Ni) between a checkerboard antiferromagnetic and ferromagnetic state for La-based layered nickelates with $n=3$, 4, 5, 6, and $\infty$ within GGA+$U$ as a function of $U$ using the AMF double counting scheme. A positive energy difference indicates the FM state is energetically favored, while a negative difference indicates the AFM state is favored. A qualitative description of the ground state (GS) of the system at each value of $U$ is provided. Finally, the Ni magnetic moments for both AFM and FM spin configurations are given for each Ni in each layer. In the 112 material, in spite of the moments remaining below 1$\mu_{B}$ within AMF at all $U$s, already for a low $U$ value, a $d_z^2$ band is pinned at the Fermi level like in the HS state described in Refs. \onlinecite{Kapeghian2020, pickett3} obtained within FLL. }
\label{tab:GGAU_AMF}
\centering
\begin{tabular*}{2\columnwidth}{c@{\extracolsep{\fill}}clccc}
    \hline
    \hline
    $U$ (eV)  &  $E_{\text{AFM}}-E_{\text{FM}}$ (meV/Ni) & GS  & NiO$_{2}$ Layer & AFM Moments ($\mu_{B}$) & FM Moments ($\mu_{B}$) \\
    \hline
    \multicolumn{6}{c}{{ \bf La438}} \\
    \hline
    1.5 & 17.4 & LS-FM  & inner &  0.86/-0.86 & 0.67 \\
        &     & metal & outer &  0.70/-0.69 & 0.66 \\
\hline
    2.7  & 38.6 & LS-FM        & inner & 1.2/-1.2 & 0.68  \\
         &      & metal & outer & 1.1/-1.1 & 0.67  \\
\hline
    4.0 & 16.0 & LS-FM   & inner & 1.3/-1.3 & 0.68 \\
        &      & metal   & outer & 1.1/-1.1 & 0.67 \\
\hline
    5.5 & 49.4 & LS-FM & inner & 1.4/-1.4 & 0.67 \\
        &   & metal    & outer & 1.2/-1.2 & 0.66 \\
\hline 
    \multicolumn{6}{c}{{ \bf La5410} }\\
    \hline
    1.5 & 13.5 & LS-FM & inner & 0.72/-0.72 & 0.74 \\
        &      & metal & outer &  0.49/-0.50 & 0.71  \\
    \hline
    2.7 & 80.9 & LS-FM & inner & 1.1/-1.1 & 0.75 \\
        &      & metal & outer & 0.57/-0.57 & 0.72 \\
    \hline 
    4.0 & 42.6 & LS-FM & inner & 1.2/-1.2 & 0.74 \\
        &     &  metal     & outer & 0.47/-0.45 & 0.71\\
    \hline 
    5.5 & 25.3 & LS-FM & inner & 1.3/-1.3 & 0.73\\
        &      & metal & outer & 0.38/-0.38 & 0.69\\
    \hline 
    \multicolumn{6}{c}{{ \bf La6512} }\\
    \hline
    1.5 & 43.5 & LS-FM & inner & 0.66/-0.66 & 0.78  \\
        &      & metal & next-to-inner & 0.62/-0.62 & 0.76 \\
        &      &       & outer & 0.52/-0.52 & 0.73 \\
    \hline
    2.7 & 133.3 & LS-FM & inner & 0.73/-0.73 & 0.78\\
        &       & metal & next-to-inner & 0.68/-0.68 & 0.77\\
        &       &       & outer         & 0.59/-0.59 & 0.73\\
    \hline 
    4.0 & 195.4 & LS-FM & inner & 0.75/-0.75 & 0.78\\
        &       & metal & next-to-inner & 0.72/-0.72 & 0.76\\
        &       &       & outer         & 0.62/-0.62 & 0.72\\
    \hline
    5.5 & 34.9 & LS-FM & inner & 1.2/-1.2 & 0.75\\
        &      & metal & next-to-inner & 0.30/-0.30 & 0.74\\
        &       &       & outer         & 1.1/-1.1 & 0.71\\
    \hline
    \multicolumn{6}{c}{{ \bf La7614} }\\
    \hline
    1.5 & 12.9 & LS-FM & inner & 0.78/-0.78 & 0.81\\ 
        &      & metal & next-to-inner & 0.62/-0.63 & 0.78 \\
        &       &      & outer & 0.56/-0.57 & 0.74\\
    \hline
    2.7 & 89.82 & LS-FM & inner & 1.0/-1.0 & 0.81\\
        &       & metal & next-to-inner & 0.56/-0.56 & 0.78\\
        &       &       & outer         & 0.89/-0.89 & 0.74 \\
    \hline
    4.0 & 81.9 & LS-FM & inner & 1.1/-1.1 & 0.79\\
        &       & metal & next-to-inner & 0.44/-0.44 & 0.78\\
        &       &       & outer         & 1.1/-1.1 & 0.73\\
    \hline 
    5.5 & 27.4 & LS-FM & inner & 1.2/-1.2 & 0.77\\
        &       & metal & next-to-inner & 0.33/-0.33 & 0.75\\
        &       &       & outer         & 1.1/-1.1 & 0.71\\
    \hline
  \multicolumn{6}{c}{{ \bf La112}}\\
  \hline
    1.5 & -122.4 & LS-AFM  &  & 0.79/-0.79 & 0.85 \\
        &        & metal  &    &            &  \\
    \hline
    2.7 & -86.9 & LS-AFM  &  & 0.83/-0.83 & 0.85\\
        &       & metal  &    &            &      \\
    \hline
    4.0 & -63.0 & LS-AFM &  & 0.84/-0.84 & 0.84 \\
        &       & metal  &    &            &   \\
    \hline
    5.5 & -62.3 & LS-AFM  &   & 0.70/-0.70 & 0.79 \\
        &       &   metal     &    &    & \\
    \hline  
    \hline
\end{tabular*}
\end{table*}
\endgroup

\section{LS Band Structures}
\label{sec:fm_bs}
\begin{figure*}
\includegraphics[width = 2\columnwidth]{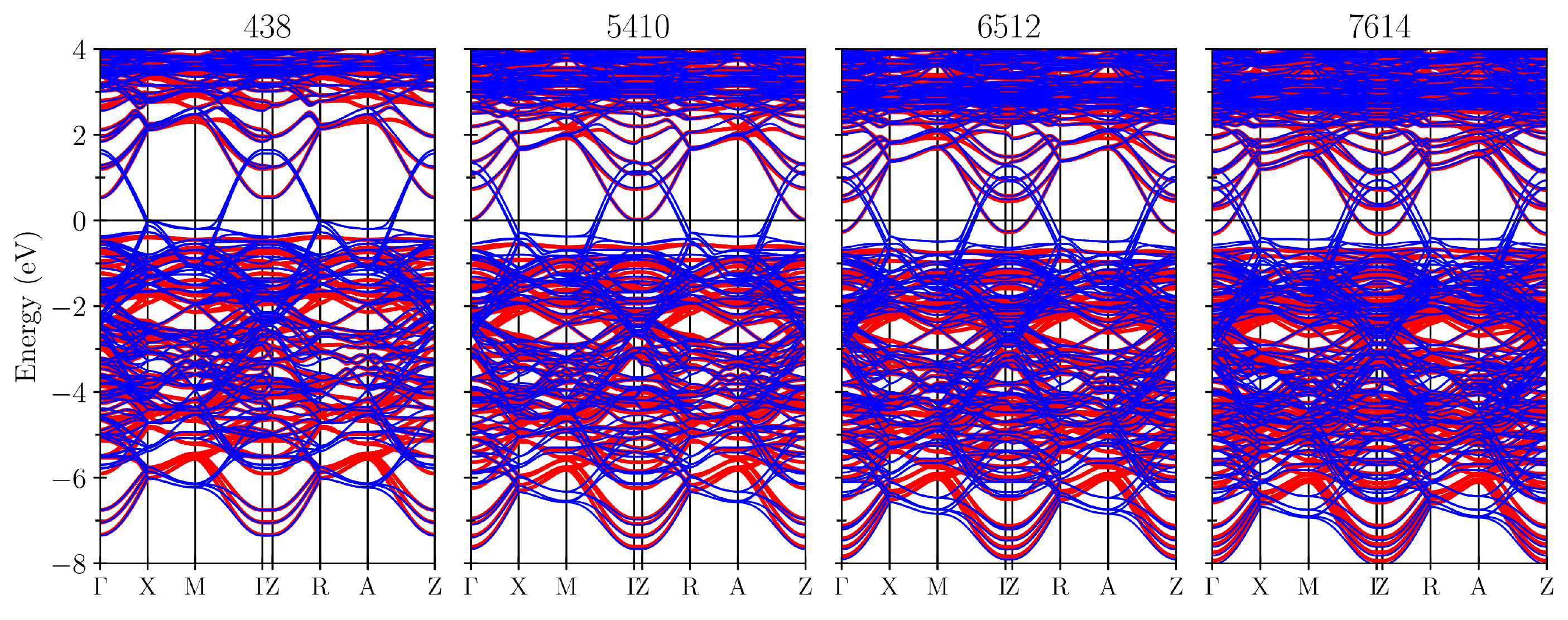}
\caption{Band structures of the $n=3-6$ phases (left to right) from GGA+$U$ (AMF) calculations within the LS-FM ground state at $U= 5.5$ eV. Blue (red) shows majority (minority) spin channels.}
\label{fig:fm_afm_bs}
\end{figure*}

The band structures from our GGA+$U$ calculations for the LS-FM ground state of $n=3-6$ layered nickelates within AMF for $U =5.5$ eV are shown in Fig. \ref{fig:fm_afm_bs}. The majority (minority) spin channels are shown in blue (red). The evolution of the electronic structure as $n$ increases repeats the same trends observed in the NM band structures shown in the main text. The La-$5d$ bands begin playing a role in the La5410 ($n=4$) compound and gradually increase their relevance in the La6512 ($n=5$) and La7614 ($n=6$) -- this effect is particularly noticeable at $\Gamma$.

\end{document}